\documentclass[sigconf]{acmart}
\AtBeginDocument{%
  \providecommand\BibTeX{{%
    \normalfont B\kern-0.5em{\scshape i\kern-0.25em b}\kern-0.8em\TeX}}}

\copyrightyear{2023}
\acmYear{2023}
\setcopyright{rightsretained}
\acmConference[ITiCSE 2023]{Proceedings of the 2023 Conference on Innovation
and Technology in Computer Science Education V. 1}{July 8--12, 2023}{Turku,
Finland}
\acmBooktitle{Proceedings of the 2023 Conference on Innovation and
Technology in Computer Science Education V. 1 (ITiCSE 2023), July 8--12, 2023,
Turku, Finland}\acmDOI{10.1145/3587102.3588785}
\acmISBN{979-8-4007-0138-2/23/07}

\usepackage{todonotes}
\usepackage{balance}
\begin{document}

%\title{Comparing the Quality of Code Explanations Created by Students and Large Language Models}
%\title{Assessing the Quality of AI-Generated Code Explanations for Learning by Example}
%\title{Comparing the Quality of Student- and Large Language Model-Created Code Explanations}
\title{Comparing Code Explanations Created by\\Students and Large Language Models}


\author{Juho Leinonen}
\affiliation{
  \institution{University of Auckland}
  \city{Auckland}
  \country{New Zealand}}
\email{juho.leinonen@auckland.ac.nz}
\orcid{0000-0001-6829-9449}

\author{Paul Denny}
\affiliation{
  \institution{University of Auckland}
  \city{Auckland}
  \country{New Zealand}}
\email{paul@cs.auckland.ac.nz}
\orcid{0000-0002-5150-9806}

\author{Stephen	MacNeil}
\affiliation{
  \institution{Temple University}
  \city{Philadelphia}
  \state{PA}
  \country{United States}}
\email{stephen.macneil@temple.edu}
\orcid{0000-0003-2781-6619}

\author{Sami Sarsa}
\affiliation{
  \institution{Aalto University}
  \city{Espoo}
  \country{Finland}}
\email{sami.sarsa@aalto.fi}
\orcid{0000-0002-7277-9282}

\author{Seth Bernstein}
\affiliation{
  \institution{Temple University}
  \city{Philadelphia}
  \state{PA}
  \country{United States}}
\email{seth.bernstein@temple.edu}
\orcid{0000-0001-5767-1057}

\author{Joanne Kim}
\affiliation{
  \institution{Temple University}
  \city{Philadelphia}
  \state{PA}
  \country{United States}}
\email{joanne.kim@temple.edu}
\orcid{0000-0001-7646-2373}

\author{Andrew	Tran}
\affiliation{
  \institution{Temple University}
  \city{Philadelphia}
  \state{PA}
  \country{United States}}
\email{andrew.tran10@temple.edu}
\orcid{0000-0002-0094-1113}

\author{Arto Hellas}
\affiliation{
  \institution{Aalto University}
  \city{Espoo}
  \country{Finland}}
\email{arto.hellas@aalto.fi}
\orcid{0000-0001-6502-209X}

\renewcommand{\shortauthors}{Leinonen, et al.}

\fancyhead{}

\begin{abstract}
Reasoning about code and explaining its purpose are fundamental skills for computer scientists. There has been extensive research in the field of computing education on the relationship between a student's ability to explain code and other skills such as writing and tracing code.  In particular, the ability to describe at a high-level of abstraction how code will behave over all possible inputs correlates strongly with code writing skills. However, developing the expertise to comprehend and explain code accurately and succinctly is a challenge for many students.  Existing pedagogical approaches that scaffold the ability to explain code, such as producing exemplar code explanations on demand, do not currently scale well to large classrooms.  The recent emergence of powerful large language models (LLMs) may offer a solution.  In this paper, we explore the potential of LLMs in generating explanations that can serve as examples to scaffold students' ability to understand and explain code.  To evaluate LLM-created explanations, we compare them with explanations created by students in a large course ($n \approx 1000$) with respect to accuracy, understandability and length. We find that LLM-created explanations, which can be produced automatically on demand, are rated as being significantly easier to understand and more accurate summaries of code than student-created explanations.  We discuss the significance of this finding, and suggest how such models can be incorporated into introductory programming education.

\end{abstract}






\maketitle

\section{Introduction}

The ability to understand and explain code is an important skill for computer science students to develop~\cite{wang2020step, cunningham2022bringing, murphy2012ability}.  Prior computing education research tends to suggest that proficiency at explaining code develops for novices after lower-level code tracing skills and is a prerequisite for higher-level code writing skills \cite{lister2009further, sheard2008going}.  After graduating, students will also be expected to explain their code to hiring managers during job interviews,  explain code to their peers as they onboard new team members, and explain code to themselves when they first start working with a new code base.   However, students struggle to explain their own code and the ability to explain code is a difficult skill to develop~\cite{Lehtinen2021students, simon2011explaining}. These challenges are further compounded by the fact that the ability to explain code is not always explicitly included as a learning objective in CS courses.

Learning by example is an effective pedagogical technique, often employed in programming education \cite{zhi2019exploring, abdulrahman2014learning}.  However, generating good examples for certain kinds of resources, such as code explanations, can be time-consuming for instructors. While learnersourcing techniques could be used to generate code explanations efficiently by directly involving students in their creation~\cite{ pirttinen2018crowdsourcing, leinonen2020crowdsourcing}, there are known issues relating to quality when learning content is sourced from students~\cite{denny2009quality, abdi2021evaluating}. In search of a remedy to this problem, researchers have explored the potential of `robosourcing' (i.e., using AI-based generators to create content or scaffold content creation by humans) learning materials~\cite{denny2022robosourcing, sarsa2022automatic}, including code explanations~\cite{macneil2022generating, macneil2023experiences}. At this stage, very little is known about how the quality of AI-generated code explanations compare with code explanations created by instructors or by students, and whether they could be used as a replacement for either.

We compare the quality of learnersourced code explanations against robosourced code explanations to examine the potential of large language models (LLMs) in generating explanations for students to use as examples for learning.  We used LLMs to create code explanations of three functions, and we asked students to create explanations of the same functions.  We then measured students' perceptions of the quality of explanations from both sources.  To aid in the interpretation of our results, we elicit from students the characteristics of a code explanation that they find most useful. The following two research questions have guided this work:

\begin{enumerate}
    \item[RQ1] To what extent do code explanations created by students and LLMs differ in accuracy, length, and understandability?
    \item[RQ2] What aspects of code explanations do students value?
\end{enumerate}

Our results show that the code explanations generated by LLMs and by students are equivalent in terms of ideal length, but that the LLM-generated explanations are perceived as more accurate and easier to understand.  Although there are benefits for students in being actively involved in producing their own explanations, we conclude that LLM-generated explanations can serve as good examples for students in early learn-by-example contexts and can be a viable alternative for learnersourced code explanations.

\section{Related Work}

\subsection{Code Comprehension}

Code comprehension skills are important for helping programming students understand the logic and functionality behind code snippets~\cite{10.1145/2325296.2325319}.
Programmers can employ various code comprehension strategies that give them flexibility in the ways they comprehend programming concepts ~\cite{402076}. Some strategies include trace execution~\cite{5441291}, explanations~\cite{10.1145/3274400}, and notional machines~\cite{guo2013online}. These strategies take time and vary in effectiveness between students ~\cite{10.1145/3387904.3389283}.
Regardless, students may face roadblocks, including logical errors~\cite{10.1145/3160489.3160493} and syntactical errors~\cite{10.1145/2325296.2325318} when trying to understand code.

Top-down and bottom-up learning are two approaches to learning that focus on the big picture and the details, respectively~\cite{Wu2019/11}. Top-down learning starts with the high-level concept and works its way down to the specifics, while bottom-up learning begins with the details and gradually works up to the high-level ~\cite{inbook}. Both approaches can be useful when teaching complex topics, as they provide a way for learners to understand the whole concept by understanding its parts. In computer science and programming, these two approaches can be used to help learners understand the fundamentals of coding and programming~\cite{10.1145/199688.199696}.

\subsection{Pedagogical Benefits of Code Explanations}

Explanations are vital teaching resources for students. Explanations help students develop their understanding of how a code snippet executes~\cite{marwan2019impact}, which can help students improve their reasoning about writing their own code~\cite{murphy2012ability}. They also reduce stress by breaking down complex concepts~\cite{griffin2016learning}.

Early approaches for code explanation, such as the BRACElet project, provided students with `explain-in-plain-English' type questions to encourage students to explain the purpose of their code at a higher level of abstraction~\cite{whalley2006australasian}. This process of explaining one's own code provided both short and long-term learning benefits for students~\cite{vihavainen2015benefits, murphy2012ability}. In large classrooms, the process of explaining code can also be a collaborative activity where peers explain code to each other. This process can be more informal, such as in the case of pair programming when students explain their code and their thought process to a partner as they write their code~\cite{hanks2011pair}.

Even though explaining code is an important skill and previous work has explored code explanation tasks, students are rarely exposed to example code explanations, especially ones created by their peers. Having easily available example code explanations could help expose students to code explanations, which could support learning to explain their own code. Having the instructor create such explanations is a time-consuming task. In big classrooms, it would be hard to find the time to provide personalized explanations for students~\cite{ullah2018effect}. Thus, studying if such explanations could be created at scale with the help of LLMs is a relevant research topic.

\subsection{Large Language Models in CS Education}

The recent emergence of AI-based code generation models has sparked considerable interest within the field of computing education research~\cite{becker2023programming}. Initial studies in this area have primarily focused on evaluating the performance of these models when solving programming problems commonly encountered in introductory courses. A seminal study in this field, entitled ``The Robots are Coming''~\cite{finnieansley2022robots}, utilized the Codex model and a private repository of programming problems drawn from high-stakes summative assessments. 
The results of the study indicated that the solutions generated by Codex scored approximately 80\% on the assessments, surpassing the performance of three-quarters of students when compared to historical course data.  Similar work involving a public dataset of programming problems found that Codex produced correct solutions on its first attempt approximately half of the time, increasing to 80\% when repeated attempts and minor adjustments to the input prompt were allowed ~\cite{denny2022conversing}.

In addition to evaluating performance, a complementary body of research has investigated the potential of AI-based code-generation models to generate learning resources. For example, Sarsa et al. explored various prompts and approaches for using the Codex model to generate code explanations and programming exercises, finding that it frequently produced novel and high-quality resources~\cite{sarsa2022automatic}. However, their evaluation was conducted solely by experts and did not involve the use of resources by students in a practical setting.  MacNeil et al. used the GPT-3 model to generate explanations of short code fragments which then were presented to students in an online e-book alongside the corresponding code~\cite{macneil2023experiences}.  Although their evaluation was conducted on a small scale with approximately 50 participants, students found the explanations to be useful when they chose to engage with them. However, as the authors noted, this engagement was lower than anticipated,  and the students were not involved in the creation of either the code examples or the accompanying explanations.

The current study makes a unique contribution by directly comparing code explanations generated by students with those generated by AI models. While prior research has demonstrated that LLMs can produce explanations of code that are deemed high-quality by both experts and novices, this is the first study to investigate how students evaluate code explanations generated by their peers in comparison to those generated by AI models.

\section{Method}

\begin{figure*}[h]
\centering
\includegraphics[width=\textwidth]{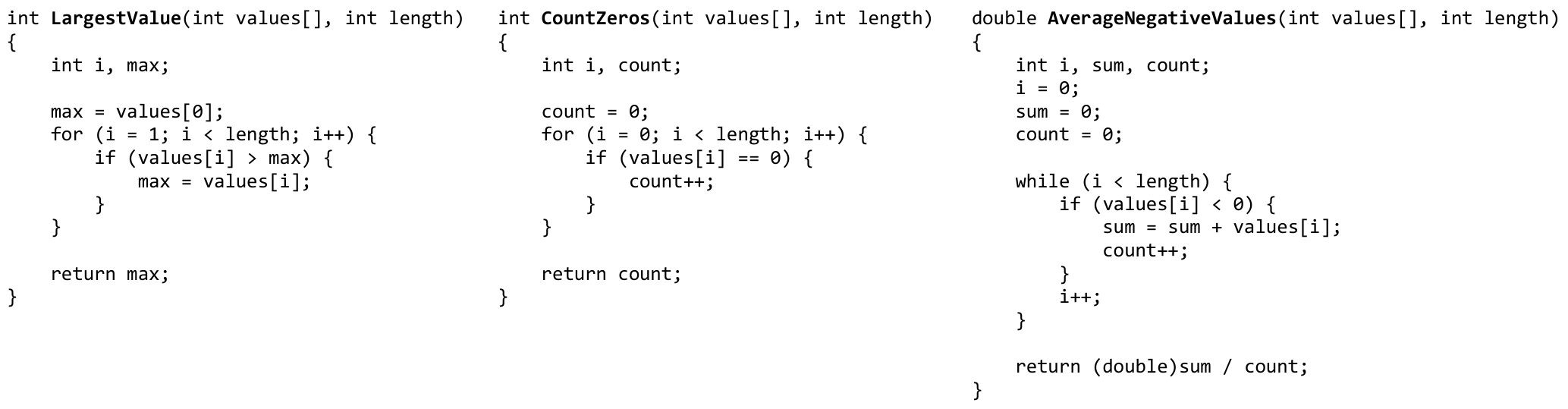}
\caption{The three function definitions, as presented to students in Lab A.  Students were asked to construct a short description of the intended purpose of each function.  \label{fig:ThreeFunctions}}
\end{figure*}

\subsection{Context and Data}

Our data for this study was collected in a first-year programming course at The University of Auckland.  Approximately 1000 students were enrolled in the course in 2022 when our study was conducted.

\subsubsection{Data collection}

The data was collected during two separate lab sessions, each of which ran over a one-week period.  At the time of the first lab, when the data collection began, the course had covered the concepts of arithmetic, types, functions, loops and arrays in the C programming language. The data collection followed the ethical guidelines of the university.

During the first lab, Lab A, students were shown three function definitions and were asked to summarize and explain the intended purpose of each function.  During the second lab, Lab B, which was conducted two weeks after the first, students were shown a random sample of four code explanations for the functions in Lab A.  Some of these code explanations were selected from the explanations generated by students during Lab A, and some were generated by the large language model GPT-3~\cite{brown2020language}. Students were asked to rate the explanations with respect to accuracy, understandability and length.  At the end of Lab B, students were invited to provide an open-response answer to the following question: ``Now that you have created, and read, lots of code explanations, answer the following question about what you believe are the most useful characteristics of a good code explanation: What is it about a code explanation that makes it useful for you?''

Figure \ref{fig:ThreeFunctions} lists the three functions that were shown to students in Lab A.  Each function includes a single loop that processes the elements of an array that is passed as input to the function, and has a name that is representative of the algorithm being implemented.  For each of the three functions, students were asked to summarize and explain the intended purpose of the function.  Specifically, they were asked to: ``look at the name of the function, the names of the variables being used, and the algorithm the function implements and come up with a short description of what you believe is the intended purpose of the function''.

\subsubsection{Data sampling}

\begin{figure}[h]
\centering
\includegraphics[width=\columnwidth]{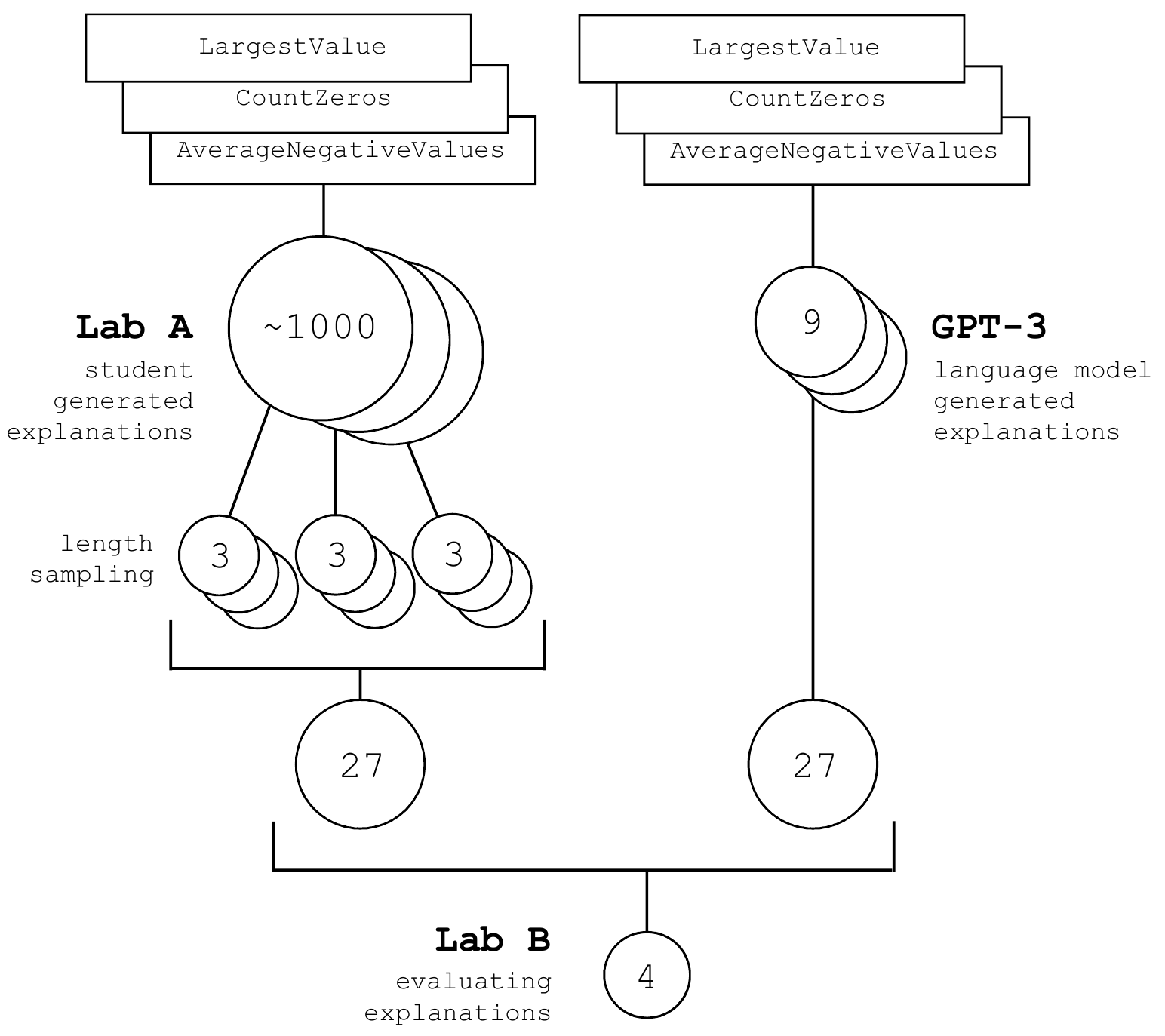}
\caption{Overview of the generation and sampling of code explanations.  In Lab B, each student was allocated four code explanations to evaluate, selected at random from a pool of 54 code explanations (half of which were generated by students in Lab A, and half of which were generated by GPT-3.)  \label{fig:schematic}}
\end{figure}

Figure \ref{fig:schematic} provides an overview of the process used to sample the code explanations used in Lab B.  Students who participated in generating code explanations in Lab A submitted 963 explanations for each of the three functions.  For each of the functions, we stratified the code explanations into three categories based on their word length: 10th percentile, 10-90th percentile and 90th percentile.    From each of these three categories, we randomly selected three explanations, resulting in nine explanations for each of the three functions.  To these 27 student-generated explanations, we added 27 explanations created by GPT-3, by generating nine explanations for each of the three functions.  For Lab B, each student was shown four explanations selected at random from the pool of 54 explanations.  They were asked to rate each of these with respect to the following three questions (each on a 5-point scale):

\begin{itemize}  
\item This explanation is easy to understand (5-items: Strongly disagree, Disagree, Neutral, Agree, Strongly agree)
\item This explanation is an accurate summary of the code  (5-items: Strongly disagree, Disagree, Neutral, Agree, Strongly agree)
\item This explanation is the ideal length (5-items: Much too short, A little too short, Ideal, A little too long, Much too long)
\end{itemize}

\subsubsection{Analyses}

To answer RQ1 and to quantify differences between student-created and LLM-generated code explanations, we compared student responses to the Likert-scale questions between the two sources of code explanations.

As Likert-scale response data is ordinal, we used the non-para\-met\-ric Mann--Whitney U test~\cite{mann1947test} to test for differences in Likert-scale question data between student and LLM code explanations. We tested: (1) whether there was a difference in the code explanations being easy to understand; (2) whether there was a difference in the code explanations being accurate summaries of the code; and (3) whether there was a difference in the code explanations being of ideal length. Further, we (4) studied the actual length of the code explanations to form a baseline on whether the lengths of code explanations differed between students and GPT-3, which could help interpret other findings.

Altogether, we conducted four Mann--Whitney U tests. To account for the multiple testing problem, we used Bonferroni corrected $p < 0.05 / 4$ as the threshold of statistical significance. Following the guidelines of~\cite{wasserstein2016asa} and the broader discussion in~\cite{sanders2019inferential}, we use $p$ values as only one source of evidence and outline supporting statistics including two effect sizes -- Rank-Biserial (RBC) Correlation~\cite{kerby2014simple} and Common-Language Effect Size (CLES)~\cite{mcgraw1992common} -- when presenting the results of the study.

To answer RQ2, i.e., examine what aspects of code explanations students value, we conduct a thematic analysis of 100 randomly selected student responses to the open-ended question ``What is it about a code explanation that makes it useful for you?''.

\section{Results}

\subsection{Descriptive Statistics}

Overall, a total of 954 students participated in the activity where they assessed the quality of code explanations. The averages and medians for the responses, where Likert-scale responses have been transformed to numeric values, are shown in Table~\ref{tbl:descriptive-statistics}, accompanied with the mean code explanation length for both student-created code explanations and LLM-generated code explanations.

\begin{table}[h]
\caption{Descriptive statistics of student responses on code explanation quality. The responses that were given using a Likert-scale have been transformed so that 1 corresponds to `Strongly disagree' and 5 corresponds to `Strongly agree'.\label{tbl:descriptive-statistics}}
\begin{tabular}{lrr|rr}
\toprule
{} & \multicolumn{2}{c|}{Student-generated} & \multicolumn{2}{l}{LLM-generated} \\
{} & Mean & Median & Mean & Median  \\
\midrule
Easy to understand & 3.75 & 4.0 & 4.12 & 4.0 \\
Accurate summary & 3.78 & 4.0 & 4.0 & 4.0 \\
Ideal length & 2.75 & 3.0 & 2.66 & 3.0 \\
\midrule
Length (chars) & 811 & 738 & 760 & 731 \\
\bottomrule
\end{tabular}
\end{table}

\begin{figure*}[h]
\centering
\includegraphics[width=0.75\textwidth]{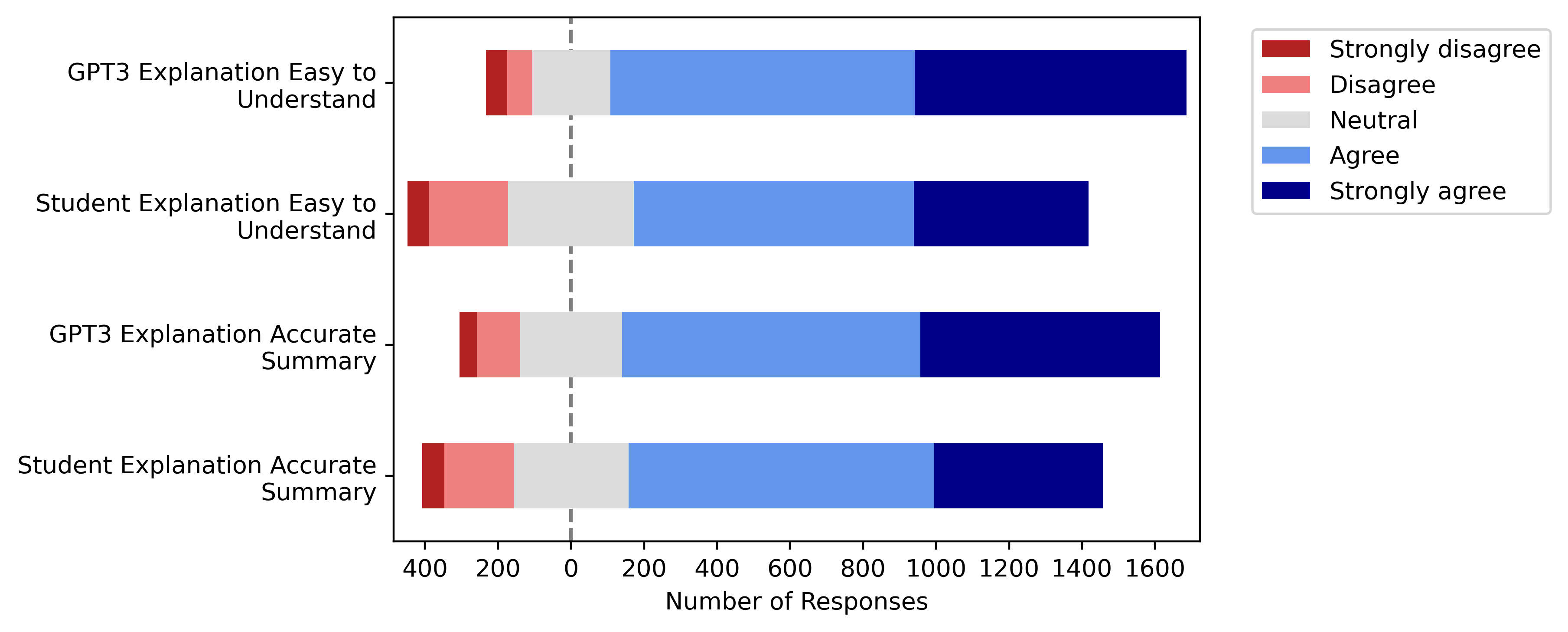}
\caption{Distribution of student responses on LLM and student-generated code explanations being easy to understand and accurate summaries of code.\label{fig:distribution-of-responses}}
\end{figure*}

Figure~\ref{fig:distribution-of-responses} further outlines the distribution of the responses, separately color coding the different responses and allowing a visual comparison of the different response values, which the numerical overview shown in Table~\ref{tbl:descriptive-statistics} complements.

\subsection{Differences in Quality of Student- and LLM-Generated Code Explanations}

Mann-Whitney U tests were conducted to study for differences between the student- and LLM-generated code explanations. We used two-sided tests, assessing for differences in the code explanations being easy to understand, accurate summaries of the shown code, and of ideal length. We further looked for differences between the actual length (in characters) of the code explanations.

The results of the statistical tests are summarized in Table~\ref{tbl:mwu-test-statistics}. Overall, we observe statistically significant differences between the student- and LLM-generated code explanations in whether they are easy to understand and in whether they are accurate summaries of the code. As per Bonferroni correction, there is no statistically significant difference in student-perceptions of whether the code explanations were of ideal length, and there is no statistically significant difference in the actual length of the code explanations.

\begin{table}[h]
\caption{Mann-Whitney U test results from two-sided comparisons in the quality of the student- and LLM authored code explanations. The $U-val$ stands for the Mann-Whitney U test U value, $p$ outlines the probability (uncorrected) that the responses come from the distribution, $RBC$ is the Rank-Biserial Correlation of the data, and $CLES$ provides the Common-Language Effect Size.\label{tbl:mwu-test-statistics}}
\begin{tabular}{lllrr}
\toprule
{} & U-val & $p$ & $RBC$ & $CLES$  \\
\midrule
Easy to understand & 1421596.0 & $3.010 \cdot 10^{-31}$ & 0.206 & 0.397 \\
Accurate summary & 1568575.5 & $2.866 \cdot 10^{-12}$ & 0.123 & 0.438 \\
Ideal length & 1860870.5 & 0.025 & -0.040 & 0.520 \\
\midrule
Length (chars) & 1779386.0 & 0.764 & 0.006 & 0.497 \\
\bottomrule
\end{tabular}
\end{table}

Overall, interpreting the common language effect size (CLES) from Table~\ref{tbl:mwu-test-statistics}, the proportion of student-generated and LLM-generated code explanation pairs where the student-generated code explanation is easier to understand is approximately 40\%, while the proportion of pairs where the LLM-generated code explanation is easier to understand is approximately 60\%. Similarly, the proportion of student-generated and LLM-generated code explanation pairs where the student-generated code explanation is a more accurate summary is approximately 44\%, while the proportion of pairs where the LLM-generated code explanation is a more accurate summary is approximately 56\%. Although these differences are statistically significant (and visible, as seen in Figure~\ref{fig:distribution-of-responses}), the effect sizes can be considered weak ($RBC < 0.3$).

\subsection{Examples of Good and Bad Explanations from GPT-3 and Students}

Our quantitative results suggest that LLM-generated explanations were more accurate and easier to understand than those generated by students. In the thematic analysis related to what students value in code explanations, we found that the LLM-generated and student-generated explanations varied in different thematic qualities. These include differences in levels of accuracy, completeness, abstraction, and expert vocabulary used. In an open-ended response, most students described a good explanation as one that “succinctly summarized the function of the code by describing what its inputs [and outputs] are,” illustrated “how the code works in coding terms but also explains what the code is trying to achieve in basic English terms,” and was “nice and simple but still has a good amount of length to describe the whole code.”

Next, to illustrate the types of explanations that GPT-3 and students created, we present four example explanations that differ in terms of student-perceived quality -- the highest- and lowest-rated LLM- and student-created explanations for the LargestValue function. The examples were ranked based on the average between accuracy and understandability.

The following LLM-generated explanation was rated the highest (understandability = 4.2, accuracy = 4.5, length = 3.5):
\begin{quote}
“This code snippet finds the largest value in an integer array. It starts by assuming that the first value in the array is the largest. It then loops through the remaining values in the array, comparing each value to the max. If a value is larger than the max, it becomes the new max. This process is repeated until all values in the array have been checked. The largest value is then returned.”
\end{quote}

The following LLM-generated explanation was rated the lowest (understandability = 4.1, accuracy = 3.7, length = 2.0):
\begin{quote}
“This code snippet returns the largest value in an array of integers.”
\end{quote}

The following student-created explanation received the highest evaluations (understandability = 4.2, accuracy = 4.5, length = 3.6):
\begin{quote}
“The function LargestValue, finds the largest integer value in the input array, and outputs that value as type int. It takes two inputs:\\

    * values - an integer array\\
    * length - the length of that array (number of elements inside the array)\\

It sets the first integer value in the array as the original ‘max’ value (largest integer in the array) and then scans through the input array (left to right), checking to find if there is a larger integer value, which if found it stores in it’s output integer variable - ‘max’, which is returned at the end of the function.”
\end{quote}

The following student-generated explanation received the lowest ratings (understandability = 3.3, accuracy = 3.3, length = 1.6):
\begin{quote}
“to find the largest value within the array”
\end{quote}

Students preferred explanations that had both information related to what the function is and how it operates being described in a line-by-line format. Many students also agreed that a good explanation explains the inputs and outputs of the code. 

Bad explanations were characterized as those that missed some details of the code while either being too long or too short. For example, an explanation may state at a high level the purpose of the code, but not go into detail about what data structures were used, or what inputs are given to the function.

Interestingly, we found that all of the LLM-generated explanations started out with the statement “This code snippet” or “The purpose of this code snippet” while the student generated explanations differed more. This was partially due to the prompting of the LLM, where it was asked to explain the purpose of ``the following code snippet''. However, most of the explanations by both students and the LLM generally followed a similar structure: function's purpose, analysis of the code, and finally the return output.

\subsection{Characterizing Code~Explanations} 

In the thematic analysis (n=100), we found that students were almost evenly split between focusing on specific (n=57) and generic (n=65) aspects of the code with some students' responses including both. When focusing on specific aspects of code students described the need for a line-by-line explanation (21\%). Students also focused on even lower-level details like the names of variables, the input and output parameters (36\%), and defining terms (8\%). Some students asked for additional aspects that were rarely included in code explanations. For example, students requested examples, templates, and the thought process behind how the code was written. 

Students commented extensively about the qualities that make a good explanation. Length was an important aspect with 40\% of the students commenting explicitly on the length of an explanation. However, there was no clear consensus about the exact length that was ideal. Instead, comments tended to focus on efficiency; conveying the most information with the fewest words. Students appeared to rate short explanations low, even when the explanation was to the point and might be something that a teacher would appreciate.
This may be partly due to such explanations giving them little or no additional information that was not already obvious in the function, e.g. the function name. Students, them being novices, likely preferred more detailed explanations since it helps them better learn and understand what is actually going on in the code.

\section{Discussion}

\subsection{Differences Between Student- and LLM-Created Code Explanations}

Github Copilot and similar tools have made code comprehension an even more important skill by shifting the focus from writing code to understanding the purpose of code, evaluating whether the code generated is appropriate, and modifying the code as needed. However, it is also possible that LLMs can not only help students to generate code, but also help them understand it by creating code explanations which can be used as code comprehension exercises.

We found that the code explanations created by GPT-3 were rated better on average in understandability and accuracy compared to code explanations created by students. This suggests that LLM-created code explanations could be used as examples on courses with the goal of supporting students in learning to read code. There were no differences in either perceived or actual length of student- and LLM-created code explanations, so the increased ratings are not due to the LLM creating longer (or shorter) explanations.

We believe that code explanations created by LLMs could be a helpful scaffolding for students who are at the stage where they can understand code explanations created by the LLM but are not yet skilled enough to create code explanations of their own. LLM-created code explanations could also be used as examples that could help students craft code explanations of their own.

One downside mentioned in previous work is potential over-reliance on LLM support~\cite{chen2021evaluating,finnieansley2022robots}. One way to combat over-reliance on LLM-created code explanations would be to monitor student use of this type of support (e.g., giving students a limited number of tokens~\cite{nygren2019experimenting} that would be used as they request explanations from an LLM) to limit student use of, or reliance, on these tools. For example, students could get a fixed number of tokens to start with and use up tokens by requesting explanations -- and then earn tokens by writing their own hand-crafted code explanations. 

\subsection{What Do Students Value in Code Explanations?}

We found in our thematic analysis that students expressed a preference for line-by-line explanations. This is also the type of explanation that LLMs seem to be best at creating~\cite{sarsa2022automatic}. This finding was somewhat surprising as prior work on `explain-in-plain-English' code explanation tasks has typically rated `relational' responses -- short, abstract descriptions of the purpose of the code -- higher than `multi-structural' -- line-by-line -- responses. This suggests that there might be a mismatch between instructor and student opinions on what makes a good explanation. It might even be that some prior work has ``unfairly'' rated student multi-structural explanations lower since students might have possibly been able to produce the more abstract relational explanations, but were thinking longer, more detailed explanations are ``better'' and thus produced those types of explanations.

In the thematic analysis, we also observed that the LLM-created explanations closely followed a standard format. It is possible that showing students LLM-created explanations could help them adopt a standard format for their own explanations, which would possibly help make better explanations. This would be similar to prior work that has shown that templates can help designers frame better problems~\cite{macneil2021framing} and writers write better emails~\cite{hui2018introassist}.

\subsection{Limitations}

There are limitations to our work, which we outline here. First, related to generalizability, the students in our study were novices. This might affect both the types of explanations they create as well as how they rate the explanations created by their peers and GPT-3. For example, prior work has found differences in how students and instructors rate learnersourced programming exercises~\cite{pirttinen2022can}. It is possible -- even likely -- that more advanced students, or e.g. instructors, could create code explanations that would be rated higher than the explanations created by GPT-3. Novices might also value different types of explanations than more advanced students: for example, it is possible that once students get more experience, they will start valuing more abstract, shorter explanations.

Related to the code being explained, we only provided students correct code in this study. An interesting avenue of future work is evaluating student and LLM performance in explaining and detecting bugs in incorrect code. The functions being explained were also relatively simple. Future work should study explanations for more varied and complex functions.

In this exploratory work, we only looked at student perceptions on the quality of the explanations. Future work should study whether there are differences in student learning when using student- and LLM-created code explanations.

We acknowledge that we analyzed the data in aggregate, i.e., some students might have only seen LLM-created explanations and some only student-created ones. We did a brief analysis of the data for students who saw two LLM-created explanations and two student-created explanations, and observed similar effects as reported in this study, and thus believe aggregating over all students is methodologically valid.

Lastly, we used the davinci-text-002 version of GPT-3. A newer version, davinci-text-003, was released in November 2022. Using the newer LLM-model would likely yield at least similar performance, if not better.

\section{Conclusion}

In this work, we presented a study where students created code explanations and then evaluated their peers' code explanations as well as code explanations created by GPT-3. We found that students rated the code explanations created by GPT-3 higher in both accuracy and understandability, even though there were no differences in the perceived or actual length of the student and LLM-created code explanations. Further, we found that students preferred detailed explanations over concise high-level explanations.

Our results suggest that LLM-created code explanations are good, and thus could be useful for students who are practicing code reading and explaining. We argue that these skills are becoming even more relevant with the advent of large language model based AI code generators such as GitHub Copilot as the role of software developers in the future will increasingly be to evaluate LLM-created source code instead of writing code from scratch.

\begin{acks}
We are grateful for the grant from the Ulla Tuominen Foundation to the first author.
\end{acks}
%\clearpage
\balance

\bibliographystyle{ACM-Reference-Format}
\bibliography{references}

%\appendix

\end{document}